\documentclass{mn2e}
\usepackage{times,psfig}

\begin{document}
\hsize=6truein

\renewcommand{\thefootnote}{\fnsymbol{footnote}}
\newcommand{\beppo}{{\it BeppoSAX} }
\newcommand{\asca}{{\it ASCA} }
\newcommand{\rosat}{{\it ROSAT} }

\title[]{\beppo observations of three distant, highly luminous
clusters of galaxies: RXJ1347-1145, Z3146 and A2390} 

\author[]
{\parbox[]{6.in} {S. Ettori, S.W. Allen and A.C. Fabian \\
\footnotesize
Institute of Astronomy, Madingley Road, Cambridge CB3 0HA 
}}                                            
% \date{Aug 2000}
\maketitle

\begin{abstract}
We present an analysis of \beppo observations of three clusters
of galaxies which are amongst the most luminous in the Universe: 
RXJ1347-1145, Zwicky~3146 and Abell~2390.
Using data from both the Low Energy (LECS) and Medium Energy (MECS)
Concentrator Spectrometers, and a joint analysis with 
the Phoswich Detection System (PDS) data above 10 keV, 
we constrain, with a relative uncertainty of between 7 and 42 per cent 
(90 per cent confidence level), the mean gas temperature 
in the three clusters.
These measurements are checked against any possible non-thermal 
contribution to the plasma emission and are shown to be robust.

We confirm that RXJ1347-1145 has a gas temperature that lies in the
range between 13.2 and 22.3 keV at the 90 per cent confidence level, and
is larger than 12.1 keV at $3 \sigma$ level. The existence of
such a hot galaxy cluster at redshift of about 0.45 implies an 
upper limit on the mean mass density in the Universe, $\Omega_{\rm m}$,
of 0.5.

Combining the \beppo estimates for gas temperature and luminosity
of the three clusters
presented in this work with \asca measurements available in the literature,
we obtain a slope of 2.7 in the $L-T$ relation once the
physical properties are corrected from the contamination
from the central cooling flows. 
% This slope is $5 \sigma$ too steep from the value of 2 predicted 
% from the gravitational collapse of gas in a cluster dark matter halo.

\end{abstract}


\section{INTRODUCTION} 

The X-ray emitting gas in clusters of galaxies cools by the emission of
energy on a time scale which depends on the temperature and
density of the intracluster medium (ICM).
Where the density is higher, in the cores of the clusters, the
cooling is more efficient and the temperature falls.
To support the outer layers of gas, a subsonic flow of gas occurs towards
the central region, producing a {\it cooling flow} that appears in X-rays
as an enhancement of the central peak in the surface brightness profile
and with a spectrum that presents multiple temperature components.

The stronger the cooling flow, 
the higher the X-ray luminosity which arises from 
both the thermal energy lost by the cooling gas and the gravitational work
done to maintain the gas at constant pressure.
Therefore, to compare global cluster gas properties with theoretical 
and numerical predictions, that, at present, cannot model the radiative cooling
in detail, we need to account fully for the effects
of cooling flows on the temperature and luminosity measurements
(Fabian et al. 1994, Allen \& Fabian 1998, Markevitch 1998).

In this paper, we present data collected with the instruments onboard
of the Italian-Dutch satellite \beppo for
three highly luminous ($L_{\rm bol}>10^{45}$ erg s$^{-1}$), 
hot ($T_{\rm gas} \ga 10$ keV),
distant ($z>0.2$) clusters of galaxies
with large ($\ga 1000 M_{\odot}$ yr$^{-1}$) mass 
deposition rates in their cores (Allen 2000): RXJ1347-1145, Zwicky~3146 
(Z3146) and Abell~2390 (A2390).
Using data from both the Low Energy (LECS) and Medium Energy (MECS)
Concetrator Spectrometer and a joint analysis with 
the Phoswich Detection System (PDS) data above 10 keV,
we show that it is possible to constrain at a high level 
of significance the hard tail of the bremsstrahlung
emission.

All quantities quoted in this paper refer to the following 
cosmological parameters: $H_0 = 50 h_{50}^{-1}$ km s$^{-1}$ Mpc$^{-1}$,
$\Omega_{\rm m} = 1$, $\Omega_{\Lambda} = 0$. 

\section{THE SAMPLE}

We describe here the characteristics of the three galaxy clusters analyzed
in the present study.

RXJ1347-1145 is a lensing cluster found in the \rosat All Sky Survey
(Schindler et al. 1995, 1997), with a slightly elongated X-ray structure 
around a very peaked central emission as it appears in the \rosat HRI image.
Analyses of the \asca dataset provide temperature estimates ranging 
from $9.3^{+1.1}_{-1.0}$ keV (Schindler et al. 1997), when a single 
temperature model is adopted, to $26.4^{+7.8}_{-12.3}$ keV (Allen 2000),
when a multi-phase gas is considered.

Z3146 is the fourth ranked luminous cluster in the \rosat band 
as reported in the Brightest Cluster Sample (BCS, Ebeling et al. 1998).
Its massive cooling flow has been inferred from the strong optical lines
and blue continuum exhibited by its central galaxy and has been studied 
with \rosat HRI (Edge et al. 1994). Allen (2000) quotes an \asca
estimate of the gas temperature corrected from the cooling flow of
$11.3^{+5.8}_{-2.7}$ keV.

A2390 is the sixth ranked luminous cluster in the \rosat band
in the northern sky (BCS, Ebeling et al. 1998), has 
been widely studied in the 
X-ray waveband (Ulmer et al. 1986, B\"ohringer et al. 1999) and is 
optically rich (Abell, Corwin \& Olowin 1989, Yee et al. 1996).
It presents evidence of strong and weak lensing (Pello et al. 1991, 
Pierre et al. 1996, Squires et al. 1996).
For A2390, B\"ohringer et al. (1999) measure $kT_{\rm gas} \sim 9$ keV when 
an isothermal model is used to fit the \asca data. 
Once they add a cooling flow component and consider a \rosat PSPC--\asca 
joint fit, the temperature rises to $11.1^{+1.5}_{-1.6}$ keV.
Using the \asca dataset only, Allen (2000)
measures $kT_{\rm gas} = 10.13^{+1.22}_{-0.99}$ keV with a single
isothermal model, that rises to $14.5^{+15.5}_{-5.2}$ keV with a
cooling flow component included in the spectral fit.

\begin{figure}
\vbox{
\psfig{figure=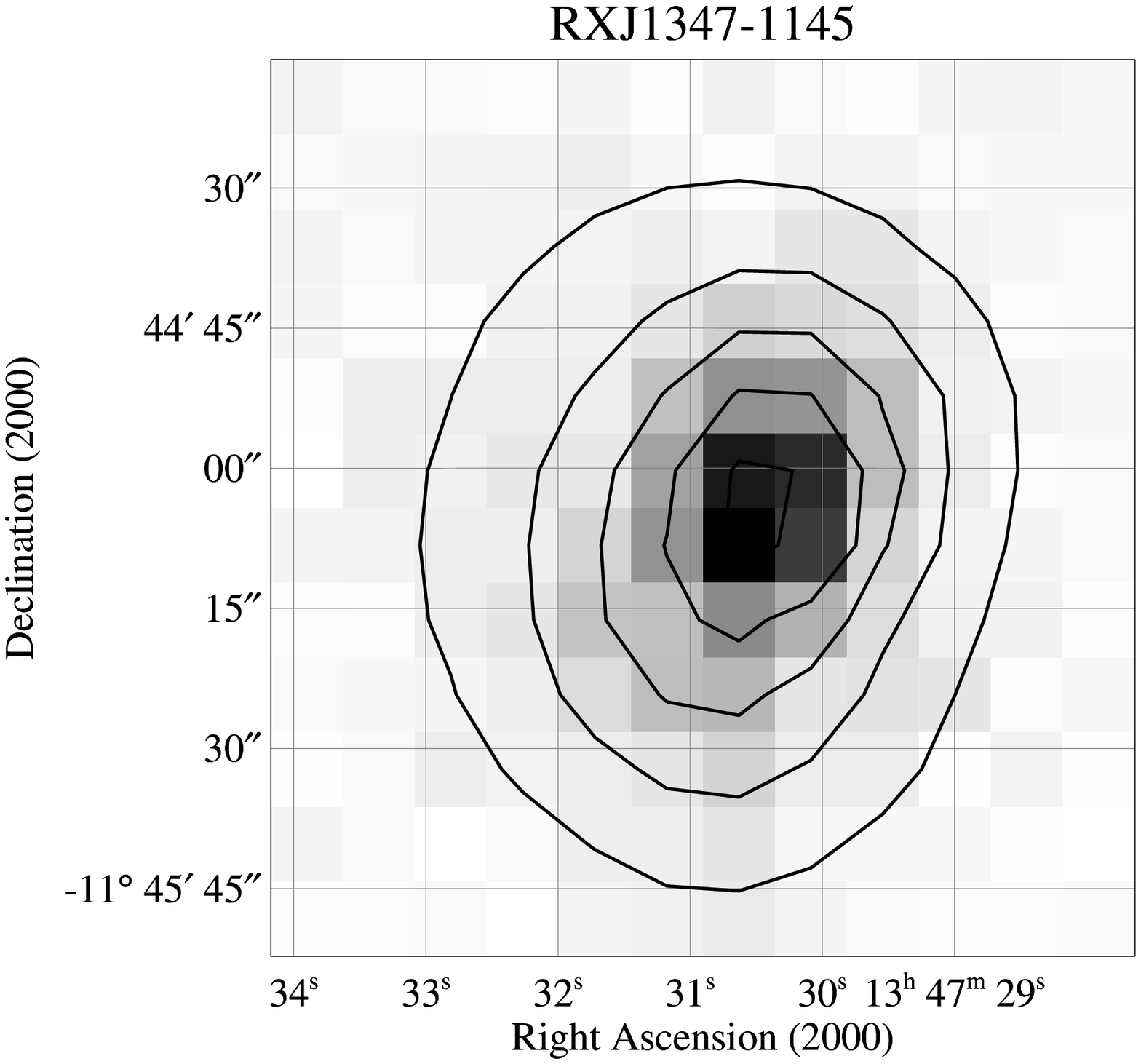,width=.4\textwidth}
\psfig{figure=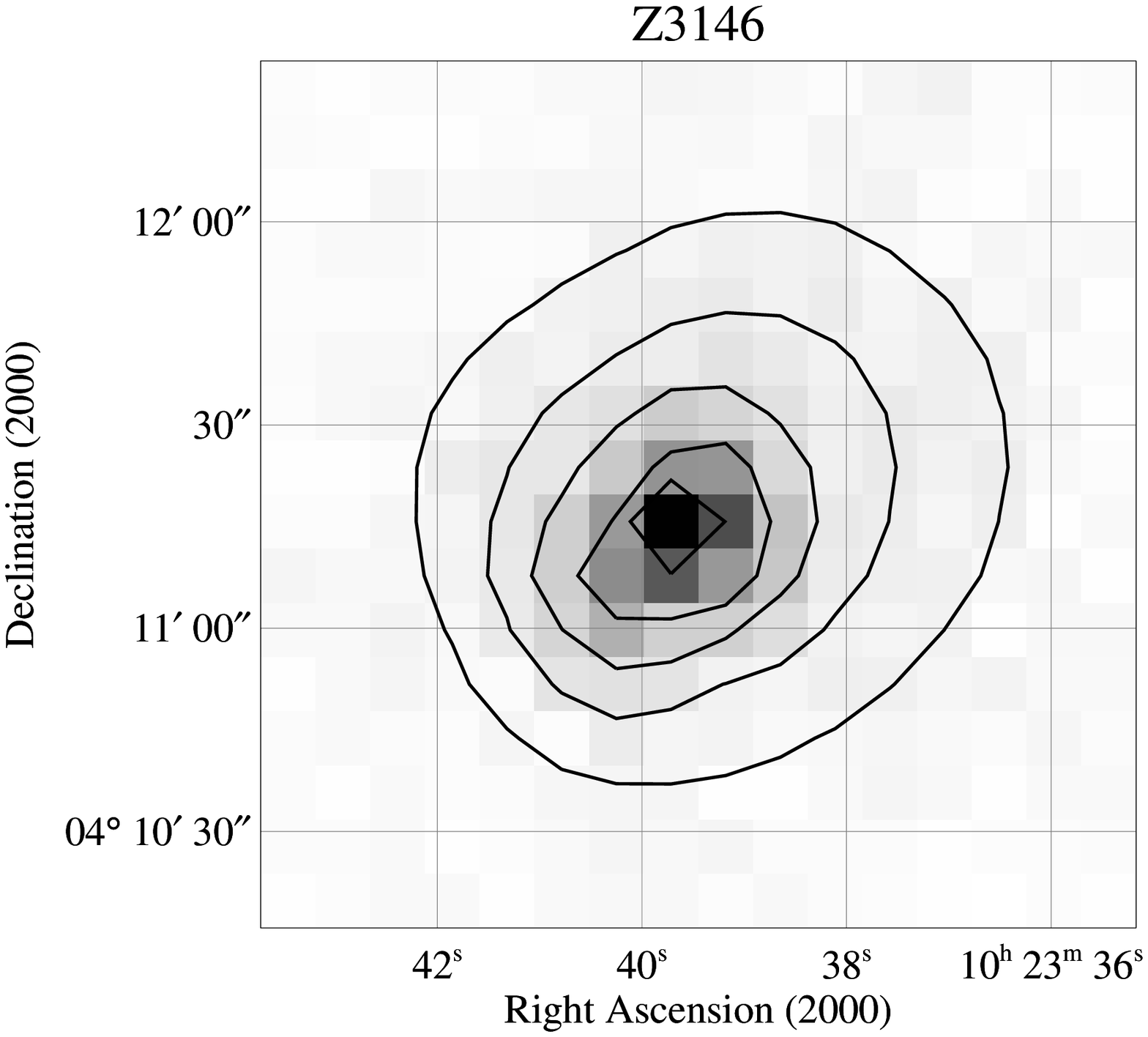,width=.4\textwidth}
\psfig{figure=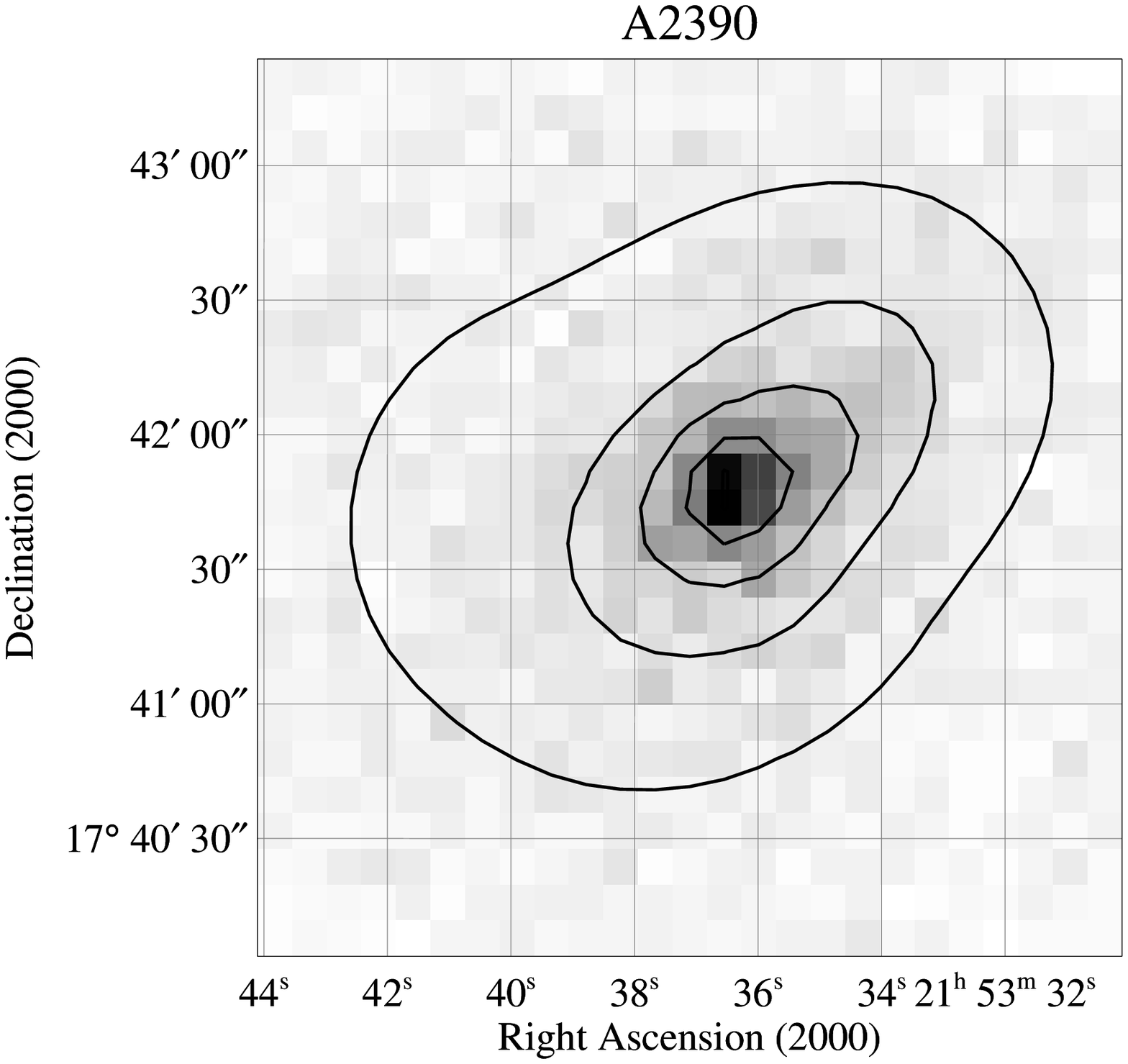,width=.4\textwidth}
}
\caption[]{ Raw \rosat HRI images with pixel-width of 5 arcsec
for the three clusters in our sample, with adaptively 
smoothed contours, obtained using a Gaussian kernel that encloses 
a number of counts equal to five times the local background,
overlaid.
} \label{fig:hri} \end{figure}

In Table~1, we quote for these three clusters the exposure 
and net count rates measured in each instrument on board \beppo, 
the Galactic absorption estimated in direction of the X-ray 
centroid determined from the respective \rosat HRI image 
(see Fig.~\ref{fig:hri}) and their redshift from literature.
Given the assumed cosmology, one arcmin corresponds to 0.408, 0.324, 
0.277 $h_{50}^{-1}$ Mpc, at the redshift of RXJ1347-1145, Z3146, 
A2390, respectively.

\begin{table*}
\caption[]{Summary of the \beppo observations of the clusters in our
sample. The Galactic absorption $N_{\rm H}$ is in unit of $10^{20}$ cm$^{-2}$
(Dickey \& Lockman 1990).
References for the redshift: [1] Schindler et al. 1995, 
[2] Allen et al. 1992, [3] Yee et al. 1996.
}
\begin{tabular}{c@{\hspace{.6em}} cc@{\hspace{.6em}} c@{\hspace{.6em}} 
c@{\hspace{.6em}} c@{\hspace{.6em}} c@{\hspace{.6em}} c@{\hspace{.6em}} 
c@{\hspace{.6em}} c@{\hspace{.6em}} c@{\hspace{.6em}}}
Cluster & RA, Dec (2000) & $z$ & $N_{\rm H}$ & date & 
\multicolumn{2}{c}{LECS (0.1--4 keV)} & \multicolumn{2}{c}{MECS (2--10.3 keV)} 
 & \multicolumn{2}{c}{PDS (13--60 keV)} \\
        & hms, deg &  &  &  & Exp. (ksec) & $10^{-2}$  cts s$^{-1}$ &
 Exp. (ksec)  & $10^{-2}$ cts s$^{-1}$ & Exp. (ksec)& $10^{-2}$ cts s$^{-1}$  \\
 \\
RXJ1347-1145 & 13 47 31.0, --11 45 11 & 0.4510 [1] & 4.8 & 2000 Jan 25 & 
28.6 & 5.3$\pm$0.1 & 73.2 & 9.9$\pm$0.1 & 36.5 & 4.0$\pm$2.6 \\
Z3146 & 10 23 40.0, +04 11 11 & 0.2906 [2] & 3.0 & 1999 Dec 07 & 
13.1 & 6.2$\pm$0.2 & 38.4 & 9.3$\pm$0.2 & 17.9 & 8.2$\pm$3.6 \\
A2390 & 21 53 36.8, +17 41 44 & 0.2279 [3] & 6.8 & 1999 May 28 & 
33.0 & 9.0$\pm$0.2 & 76.1 & 16.5$\pm$0.2 & 35.4 & 2.1$\pm$2.4 \\
\end{tabular}
\end{table*}

\section{SPECTRAL ANALYSIS}

We present the data from the three instruments onboard 
the Italian-Dutch satellite \beppo (Boella et al. 1997):
LECS (0.1--4 keV), MECS (1.8--10.5 keV) and PDS (15--200 keV).
In Table~1, we present the details of the observations.

Using {\sc Xselect} v2.0, we extract the spectra after screening 
for events without an attitude solution (generally 1 count for every 
1,000 good events) from the standard cleaned and linearized event
file of LECS and the merged event file 
collected from units 2 and 3 of the MECS.

For all three clusters, we have selected circular regions with radii 
of 8 arcmin around the source in the LECS image and 6 arcmin
in the MECS image. These regions should contain more 
than 95 per cent of the flux observed in the MECS
and about 90 per cent of that collected from the LECS, given
the standard calibration files for point sources (see below for
the calibration files adopted in the present analysis). 

Using similar regions in detector coordinates,
background spectra were extracted from the {\tt MECS23\_bkg.evt} file,
a 475 ksec exposure of blank fields at high Galactic latitude.
Following the procedure described in Fiore, Guainazzi \& Grandi (1999),
we have checked that the count rates in regions of the source field 
far away from the two calibration sources, the MECS strongback and 
any evident serendipitous source are consistent
with the count rates measured in the same regions (in detector coordinates, 
DETX and DETY) in the blank-field.

For the spectral analysis, we have made use of the calibration files released
on September 1997 from the SAX-SDC, specifically the redistribution 
matrix file {\tt lecs\_sep97.rmf} and {\tt mecs1\_sep97.rmf}, 
and the ancillary response file {\tt lecs\_130\_124\_8\_sep97.arf} 
and {\tt mecs23\_6\_sep97.arf}, for the LECS and MECS, respectively.
Moreover, we have used the {\sc SAXDAS} routine {\it lemat} 
to obtain both the RMF and ARF files for the three datasets
simulating the extend emission of the cluster with $n \sim 100$ 
point sources each with an estimated contribution to the X-ray flux 
determined by the opportunely rebinned image in RAW coordinates.
  
The counts were grouped to a minimum of 25 per bin 
in order that the $\chi^2$ statistic could be 
applied in the fitting analysis.

In the process of fitting the spectra simultaneously, we have left
free the intercalibration constant of the spectral normalization in 
the LECS with respect to that in the MECS, used as reference.
This value varies between 0.71 and 0.73 when an effective area 
appropriate for a point-source is used in fitting the LECS data.
This is in agreement with the present absolute calibration 
of the two instruments. 
When the more appropriate ARF file created with {\it lemat} is used,
the intercalibration constant is about 1 for all the best-fit results.

The PDS background-corrected spectra 
were retrieved from the on-line SDC archive and analyzed
using the redistribution matrix file {\tt pds\_256\_aug97\_fixjan98.rmf}.
We group the PDS counts according to the grouping file provided 
by SDC. The PDS vs. MECS normalization factor was fixed to 0.86.
No differences are observed between PDS spectra that do or do not
include corrections due to applying a crystal-temperature-dependent
Rise-Time threshold.

\begin{figure*}
\hbox{ 
\psfig{figure=fig2a.ps,width=.5\textwidth,angle=-90}
\psfig{figure=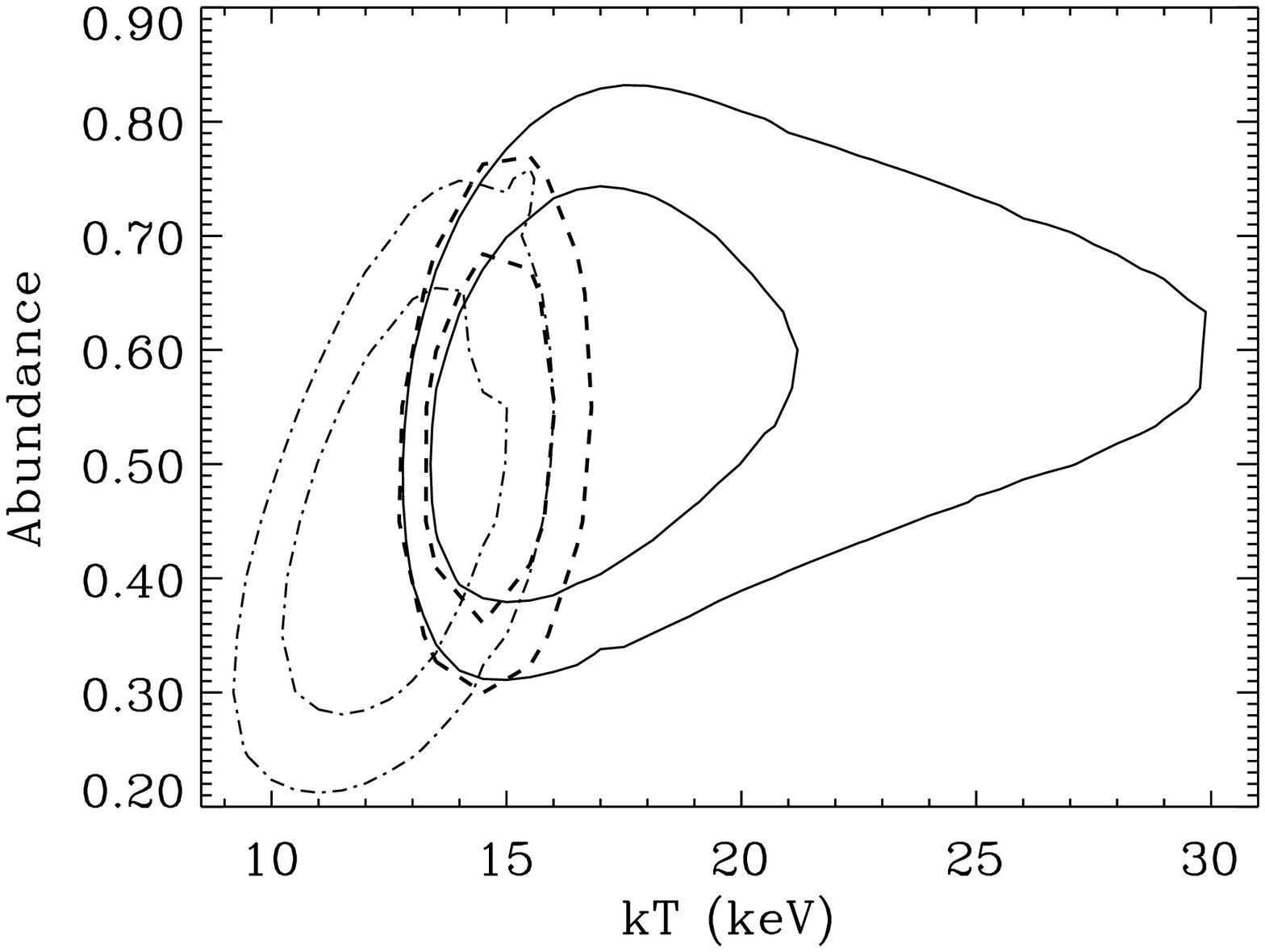,width=.5\textwidth} }
\hbox{ 
\psfig{figure=fig2c.ps,width=.5\textwidth,angle=-90}
\psfig{figure=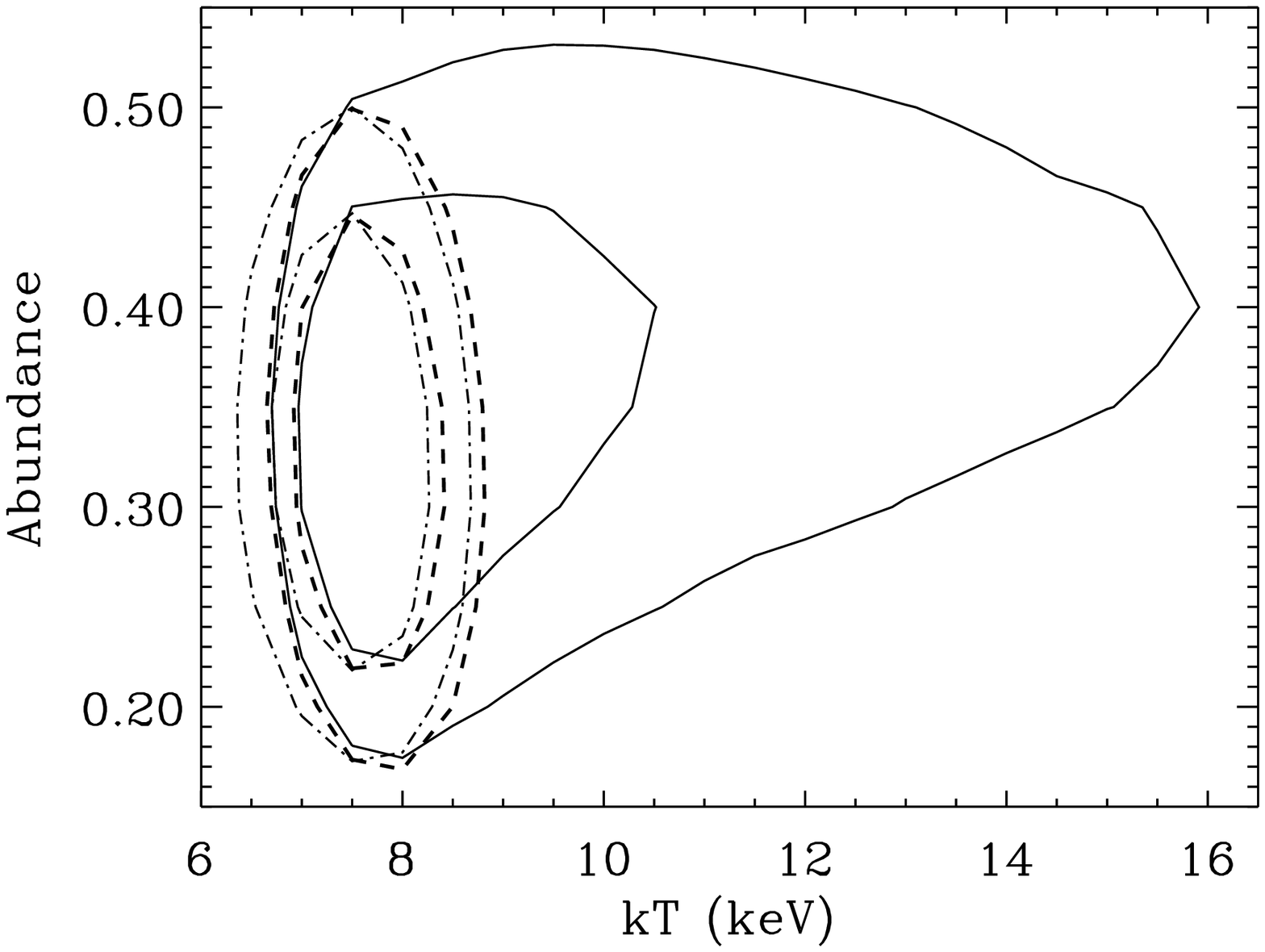,width=.5\textwidth} } 
\hbox{ 
\psfig{figure=fig2e.ps,width=.5\textwidth,angle=-90}
\psfig{figure=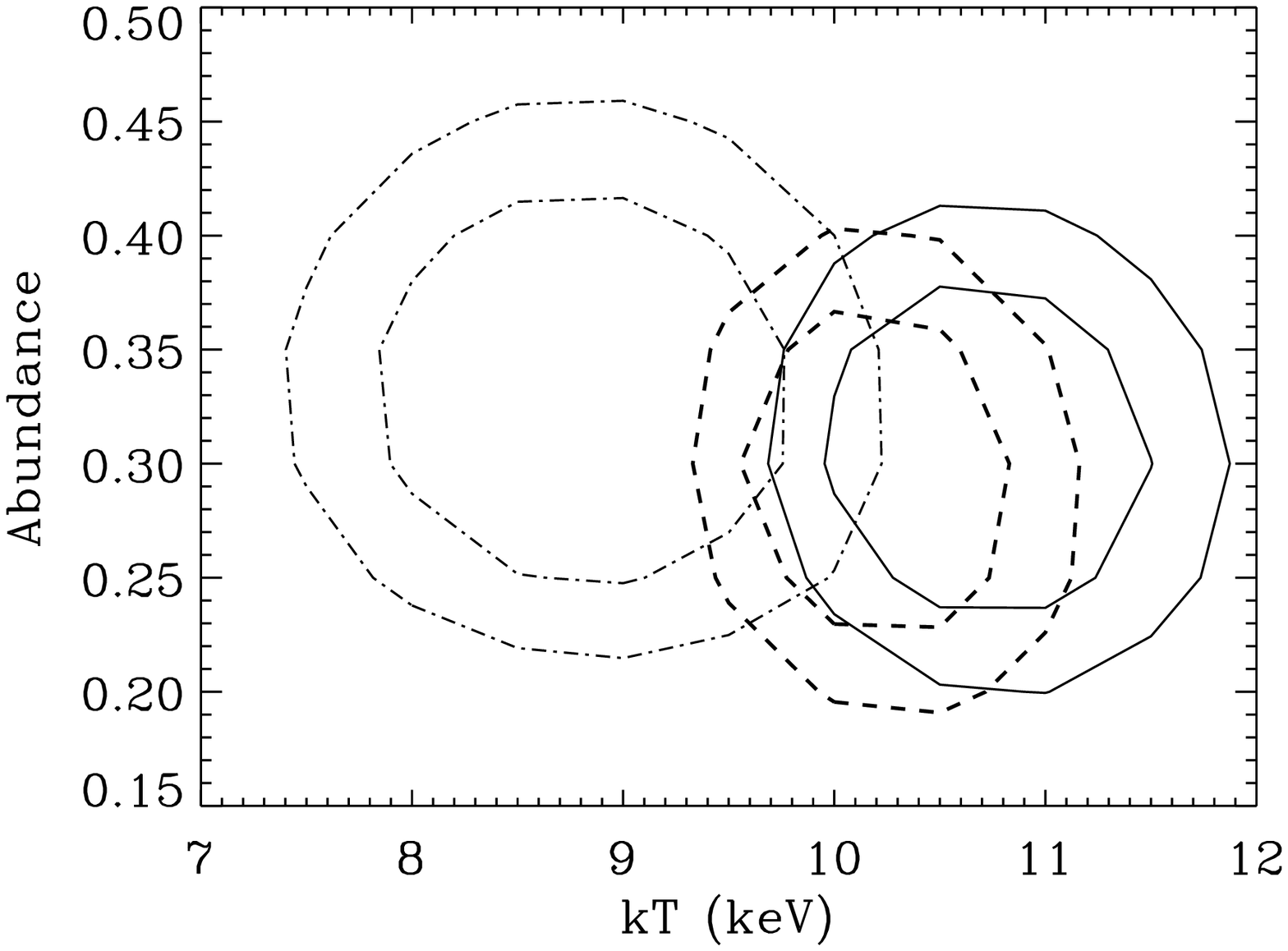,width=.5\textwidth} }
\caption[] {({\it Left}) Data and best-fit folded {\sc Mekal} model absorbed
by Galactic absorption. We plot the result of the joint fit analysis of the
LECS, MECS and PDS data and the corresponding residuals in terms of $\sigma$.
({\it Right}) Confidence contours ($\Delta \chi^2 =2.30, 4.61$
corresponding to a confidence level of 68.3 and 90 per cent, respectively,
for 2 degrees of freedom) relative to the best-fit on the two interesting 
parameters, the metal abundance and the gas temperature.
The solid line encircles the accepted values given a thermal plus cooling 
flow component; the dashed line represents the best-fit results obtained 
with a single thermal component; the dot-dashed line is obtained using a 
thermal component plus an absorbed power law (see Table~2).
} \label{fig:mecs+pds} \end{figure*}

We fit the spectra using the photoelectric absorption model 
that adopts the cross-sections calculated by Balucinska-Church 
and McCammon (1992), in front of a {\sc Mekal} emission model (Kaastra 1992, 
Liedhal et al. 1995) in {\sc Xspec} (version 11.0.1, Arnaud 1996). 
The absorption is fixed to the Galactic value quoted in Table~1 
obtained from the 0.675$^{\circ} \times$ 0.675$^{\circ}$ pixel HI map
of Dickey \& Lockman (1990).
The results of the fitting analysis are presented in Table~2 and 
shown in Figure~\ref{fig:mecs+pds}.
As comparison, we quote in Table~2 the results for a thermal fit to the 
MECS data only.

In the same Table, we also present the results obtained from fitting two 
further models that include an additional 
contribution to the single-phase plasma:
(i) a power-law that can take into account any non-thermal contamination
from an unresolved active galaxy; (ii) an intrinsically
absorbed cooling flow component which can model the emission from the
central multi-phase gas (see also discussion for the \asca data in
Allen \& Fabian 1998).
The aim in fitting these models is to check the robustness of the 
gas temperature estimates against any additional contribution to the
thermal emission. 
We note, however, that the LECS, which provides data 
in the soft X-ray part of the spectrum and is
used to constrain the intrinsic absorption and the normalization
of the cooling flow component, has an energy resolution (FWHM) of about
11 per cent at 3 keV, that is comparable to \asca Gas scintillation 
Imaging Spectrometers (GIS) but is poorer by a factor of 4 than
the Solid state Imaging Spectrometers (SIS), which also has an effective
area larger by a factor of 10 than the LECS.
Therefore, any constraint on the modeling of the low-energy spectrum
obtained through the ``1T+CF" model will be less tight of that 
provided \asca data (see Allen 2000).
In our analysis, we placed lower and upper limit on
the intrinsic cluster absorption of $10^{19}$ and $10^{22}$ cm$^{-2}$
respectively. We flag as {\it unc} the values which are 
unconstrained within these bands by our analysis.

The quoted pseudo--bolometric luminosity is estimated over the energy
range 0.01--100 keV for the ``1T+CF'' model.

\begin{table*}
\caption[]{Best-fit spectral parameters. The errors are at 90 per cent
confidence limit.
The absorption is fixed to the respective Galactic value and the redshift
to the optical estimate (cf. Table~1).
The model used in {\sc Xspec} is {\tt phabs(mekal)} (1T),
{\tt phabs(mekal+zphabs(powerlaw))} (1T+pow),
{\tt phabs(mekal+zphabs(cfmodel))} (1T+CF).
The column ``Prob" gives the null hypothesis probability that the fit 
is acceptable. The column ``$\dot{M}$-pow" quotes the deposition rate 
for the 1T+CF model, the photon index of the power law for the 1T+pow model.
The column ``$L_{\rm 2-10 keV} - L_{\rm bol}$" quotes either the luminosity 
in the 2--10 keV band for the fit on only MECS data or
the expected contribution in the same band from any power law
component (given the nominal values of the fit and upper
limit at 90 per cent confidence level for a fixed photon index of 2)
or the bolometric value calculated from the 1T+CF model.
Note that {\it unc} means unconstrained.
}
\begin{tabular}{cccccccccc}
data & Model & kT & abundance & $\chi^2_{\nu}$ (d.o.f.) & Prob &
 & $\Delta N_{\rm H}$ & $\dot{M}$-pow & $L_{\rm 2-10 keV} - L_{\rm bol}$ \\
 & & keV & $Z/Z_{\odot}$ & & & & $10^{21}$ cm$^{-2}$ & & $10^{44}$ erg s$^{-1}$\\
 \\
\multicolumn{9}{c}{{\bf RXJ1347-1145}} \\
only MECS & 1T & $14.25^{+1.79}_{-1.48}$ & $0.51^{+0.19}_{-0.17}$
& 0.989 (131) & 0.52 & & -- & -- & 83.5 \\
LECS + MECS + PDS & 1T & $14.48^{+1.76}_{-1.46}$ & $0.52^{+0.19}_{-0.18}$
& 0.993 (207) & 0.52 & & -- & -- & -- \\
 & 1T+pow & $12.64^{+2.58}_{-2.65}$ & $0.46^{+0.22}_{-0.19}$ & 0.990 (205)
 & 0.53 & & $103.4^{+475.0}_{-81.1}$ & 2 (fixed) & ($14.2, <30.2$) \\
 & 1T+CF & $15.88^{+6.47}_{-2.66}$ & $0.55^{+0.21}_{-0.19}$ & 0.999 (205)
 & 0.49 & & unc & $<1999$ & 204.0 \\ 
 \\
\multicolumn{9}{c}{{\bf Z3146}} \\
only MECS & 1T & $7.26^{+0.90}_{-0.75}$ & $0.33^{+0.13}_{-0.12}$
& 0.904 (97) & 0.74 & & -- & -- & 32.4 \\
LECS + MECS + PDS & 1T & $7.60^{+0.92}_{-0.77}$ & $0.33^{+0.13}_{-0.12}$
& 0.992 (141) & 0.51 & & -- & -- & -- \\
 & 1T+pow & $7.44^{+0.95}_{-0.85}$ & $0.33^{+0.13}_{-0.12}$ & 0.976 (139)
 & 0.57 & & $>976.8$ & 2 (fixed) & ($49.4, <106.8$) \\
 & 1T+CF & $7.75^{+3.47}_{-0.86}$ & $0.33^{+0.14}_{-0.12}$ & 1.003 (139)
 & 0.47 & & unc & $<1267$ & 69.6 \\  
 \\
\multicolumn{9}{c}{{\bf A2390}} \\
only MECS & 1T & $9.76^{+0.76}_{-0.68}$ & $0.30^{+0.08}_{-0.08}$ & 
0.991 (142) & 0.52 & & -- & -- & 34.7 \\
LECS + MECS + PDS & 1T & $10.17^{+0.77}_{-0.69}$ & $0.29^{+0.09}_{-0.08}$ &
1.074 (263) & 0.20 & & -- & -- & -- \\
 & 1T+pow & $8.79^{+1.23}_{-1.08}$ & $0.33^{+0.05}_{-0.09}$ & 1.039 (261)
 & 0.32 & & $30.8^{+18.6}_{-13.1}$ & 2 (fixed) & 
 ($7.8, <11.6$) \\
 & 1T+CF & $10.67^{+0.91}_{-0.80}$ & $0.31^{+0.08}_{-0.08}$ & 1.058 (261)
 & 0.25 & & unc & $522^{+308}_{-337}$ & 78.3 \\
 \\
\end{tabular}
\end{table*}

Generally, all of the spectral fits provide an acceptable reduced 
$\chi^2$. 
We discuss in the next subsection the results obtained from both the 
``1T" and ``1T+CF" model. 
Regarding the ``1T+pow" model, we note that
only A2390 shows a constrained contribution from a power-law component
with a luminosity less than $10^{45}$ erg s$^{-1}$ (2--10 keV) and
a photon index of $2.69^{+2.68}_{-1.31}$ consistent with the
typical value for active galaxies (e.g. Nandra et al. 1997).
Both RXJ1347-1145 and Z3146 do not show reasonable values 
to model the power-law component.
For reference, we quote in Table~2 the best-fit results
obtained for the three clusters in exam with the ``1T+pow" model and
fixing the photon index to 2 (Nandra et al. 1997).

\subsection{Comparison with \asca and \rosat results}

All of the clusters in our sample were observed with both the \asca 
(Tanaka et al. 1994) and \rosat (Tr\"umper 1983) satellites.
We discuss below the previous estimates of the gas luminosity and temperature
in comparison to the results presented in this work.

\begin{itemize}
\item {\bf RXJ1347-1145.}
Schindler et al. (1997) estimate a luminosity of $76 
\times 10^{44}$ erg s$^{-1}$ in the 2--10 keV \asca/GIS band.
This measurement is about 20 per cent lower than
the value estimated from the \asca/GIS3 for the same dataset
by Allen (2000; $L_{\rm 44, 2-10 keV} \sim 93.5$, where hereafter
$L_{\rm 44}$ is the luminosity in unit of $10^{44}$ erg s$^{-1}$).
From the fit of the MECS spectrum alone, we measure an intermediate
value between the two \asca estimates of $83.5 \times 10^{44}$ 
erg s$^{-1}$.

Schindler et al. (1997) also report a luminosity 
of $7.3$ and $\sim 21 \times 10^{45}$ erg s$^{-1}$
with \rosat/HRI in the 0.1--2.4 keV and bolometric bands, respectively.
Whereas the extrapolation of our 1T best-fit provides only
$\sim$ 70 per cent of the \rosat flux in the 0.1--2.4 keV band, 
the bolometric value is in good agreement with the result
obtained from the 1T+CF model applied to the joint analysis
of the 3 \beppo instruments and about 7 per cent lower than
the estimate in Allen (2000).

More controversial is the case for the gas temperature.
While the estimate in Allen (2000) obtained from fitting
simultaneously the 0.6--10 keV SIS and 1--10 keV GIS data
is in agreement with the best-fit value of about 14 keV within the
90 per cent confidence interval, the value of $9.3^{+1.1}_{-1.0}$
keV (90 per cent confidence level) quoted in Schindler et al. (1997)
can be excluded from the \beppo data at 99.99 per cent
confidence level. In fact, from our analysis, 
the $3 \sigma$ (99.73 per cent) lower limit is given by a 
temperature of $12.1$ keV.

We show in Fig.~\ref{fig:rxj} the residuals produced from the \beppo data
when the model of Schindler et al. (1997) is used. In particular,
it seems that most of the residuals appear at energies below $\sim 1$ keV
and above $\sim 8$ keV, in spectral region not well covered from 
the selected energy range in their \asca analysis of 0.7--9 keV.  

\begin{figure}
\psfig{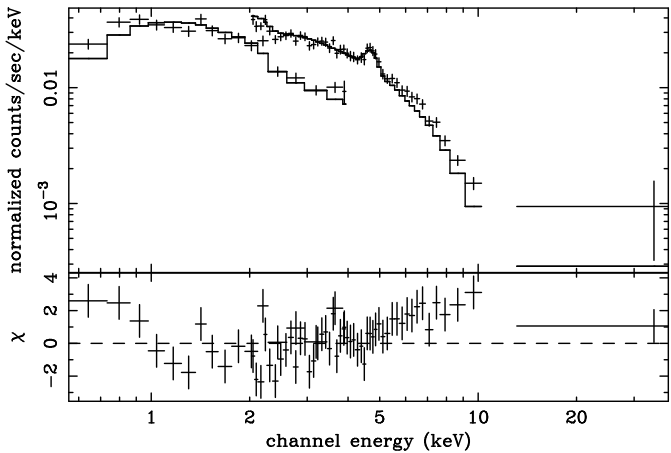}
\caption[] {
The best-fit model in second line of Table~1 in Schindler et al. (1997)
is here considered: $kT = 9.3$ keV, $N_{\rm H} = 10^{21}$ cm$^{-3}$,
$Z = 0.32 Z_{\odot}$. Only the normalization is left free to vary.
The rebinning is done for presentation purpose.
} \label{fig:rxj} \end{figure}

\item {\bf Z3146.} 
The estimated intrinsic rest-frame luminosity from the RASS 
(0.1--2.4 keV, Ebeling et al. 1998) is $26.5 \times 10^{44}$
erg s$^{-1}$. Using the same spectral model of Ebeling and 
collaborators, we calculate a $L_{\rm 44, 0.1-2.4 keV}$ of $20.0$.
Extrapolating the results of our 1T model, we measure $25.8 \times 
10^{44}$ erg s$^{-1}$. Thus, there is evidence for a slight 
mismatch in the spectrum shape between \rosat and \beppo, 
despite a general agreement in the normalization.
 
In the 2--10 keV band, the fit of the MECS spectrum provides
a luminosity of $32.4 \times 10^{44}$ erg s$^{-1}$, 13 per cent
lower than the estimate from \asca/GIS3 in Allen (2000).

Regarding the plasma temperature, our estimates are consistent
within the 90 per cent confidence level with the results 
in Allen (2000).

\item {\bf A2390.}
B\"ohringer et al. (1999) fit both \rosat/PSPC and \asca/GIS data.
They measure a luminosity of $27$ (0.1--2.4 keV) and $\sim 35$
(2--10 keV) $10^{44}$ erg s$^{-1}$.
Ebeling et al. (1998) quote an intrinsic luminosity of $21.4
\times 10^{44}$ erg s$^{-1}$ from the \rosat All Sky Survey.
Extrapolating to the 0.1--2.4 keV band the fit on our dataset, we obtain
$L_{\rm 44} = 23.7, 24.2$ using the spectral model in Ebeling et al. 
(1998) and our best-fit 1T model, respectively, in good agreement
with the \rosat estimates.

Allen (2000) quotes a 2--10 keV \asca/GIS3 luminosity of $41.0
\times 10^{44}$ erg s$^{-1}$, that is about 17 per cent higher
than both the estimate of B\"ohringer et al. (1999) and
our measure from MECS data only of $34.7 \times 10^{44}$ erg s$^{-1}$.

The gas temperature quoted in both Allen (2000) and B\"ohringer et al. 
(1999), and obtained from analysis of \asca data, are consistent 
within the 90 per cent confidence limits with the values measured
from the joint LECS-MECS-PDS fit. 
\end{itemize}

In conclusion, our results point to a general 10 per cent 
systematic disagreement among the \asca, \rosat and \beppo 
flux calibrations. 
It is worth noting that Grandi and Guainazzi show in the 
1999 report on the \beppo - \asca intercalibration using 3C273
that GIS3 provides a normalization of the continuum higher
by a factor between 9 and 15 per cent than the corresponding
value estimated in the same energy band with MECS 
for quasi-simultaneous data acquired in 1996 and 1998.

\section{A COOLING--FLOW CORRECTED L--T RELATION}

The relations between the measured (e.g. gas temperature and luminosity) 
and derivative (e.g. total gravitating mass) quantities 
determined from X-ray observations is a basic step in constraining 
cosmological parameters through the cluster mass function.

The observed values of the intracluster gas luminosity and temperature 
are correlated due to the fact that the luminosity depends on the cluster 
baryon fraction and the temperature is a direct measure of the potential 
well of the dark halo.

In fact, the luminosity, $L$, goes like $\Lambda(T) n_{\rm gas}^2 r^3 \sim
\Lambda(T) T^{3/2} (1+z)^{3/2} 
(\Omega_{\rm m}/\Omega_{\rm m}(z))^{1/2}$, where
we have considered that (i) the gas fraction ratio is constant with redshift, 
(ii) the dark matter density evolves like $(1+z)^3$, and
(iii) the equation of the hydrostatic equilibrium, $M_{\rm tot} \sim r T$,
holds.

Considering that bremsstrahlung emission is the dominant emission
process in clusters with $kT > 2$ keV, the cooling function, 
$\Lambda(T)$, is then proportional to $T^{1/2}$ and, 
consequently, the bolometric luminosity is proportional to 
the square of the temperature, $L \sim T^2$.
However, this is only valid for a sufficiently hot cluster where 
the ambient luminosity and temperature are well determined above
any contamination, like the excess in cool emission present in
the core of cooling flow clusters (e.g. Fabian et al. 1994, Allen \& 
Fabian 1998, Markevitch 1998). 

The three clusters presented in this work are among the most luminous
known and are affected by the presence of the massive central 
cooling flows. Therefore, it is worth considering
where they lie in the $L-T$ plot once we take into consideration the 
best-fit results of the ``1T+CF" model.
We add these results to a list of \asca measurements of bolometric
luminosity and temperature obtained from literature and corrected
for the contamination from cooling flows, either by including a cooling flow
model in the spectral analysis (like we have done in the present analysis;
cf. also Allen \& Fabian 1998) or excluding the central emission to compose
the analyzed spectrum (e.g. Markevitch 1998). 

\begin{figure}
\hspace*{-1.cm}
\psfig{figure=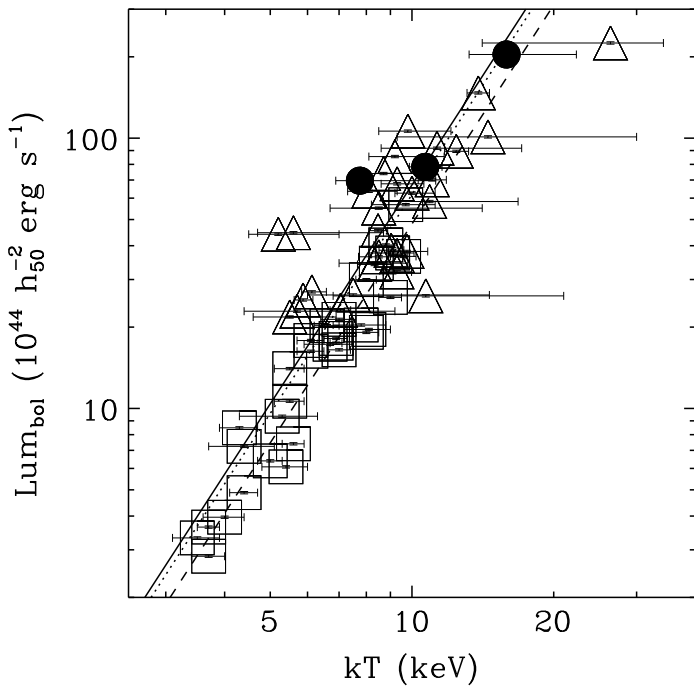,width=.6\textwidth}
\caption[]{The observed luminosity--temperature relation from \beppo and 
\asca data with the best--fit result obtained for the 64 
clusters in our sample. The {\it solid dots} represent the \beppo results
for RXJ1347-1145, Z3146 and A2390. The {\it squares} are 
from Markevitch (1998). The {\it triangles} are from Allen (2000).
All the estimated temperatures
are corrected for the presence of the cooling flow.
The error bars are 90 per cent confidence limit.
The solid, dotted, dashed lines are the best-fits obtained fixing to a common
value the slope and leaving free the normalizations for
the objects in the \beppo, Allen (2000) and Markevitch (1998)
samples, respectively. 
} \label{fig:l-t} \end{figure}

The final sample is plotted in Fig.~\ref{fig:l-t} and contains 
64 entries. Of these, 3 are in common between our sample and the one
of Allen and collaborators (A2390, Z3146, RXJ1347-1145), 
five (A478, A1795, A2029, A2142, A2319) are 
analyzed both in Allen \& Fabian (1998) and in Markevitch (1998).

The physical values of the clusters in the combined sample
range between $2.8 \times 10^{44}$ (A2657) 
and $2 \times 10^{46}$ (RXJ1347-1145) erg s$^{-1}$ in bolometric 
luminosity (the median of the distribution is $3.0 \times 10^{45}$ 
erg s$^{-1}$), between $3.5$ (A1736) and $15.9$ (RXJ1347-1145) keV 
in temperature (median: $8$ keV),
between $0.04$ (A2657) and $0.451$ (RXJ1347-1145) in redshift
(median: $0.085$).

To investigate the behavior of the $L-T$ relation, we use 
the following linear fit
\begin{equation}
\log Y = \log a +b \log X,
\label{eq:fit}
\end{equation}
where $Y = L_{\rm X} / (10^{44} h_{50}^{-2}$ erg s$^{-1})$ and
$X = T_{\rm gas} / (10 {\rm keV})$, and we use the downhill simplex method 
implemented in the {\it amoeba} routine to minimize the following $\chi^2$
merit function (Press et al. 1992):
\begin{equation}
\chi^2(a,b)_i = \sum_N \frac{(\log Y - \log a -b \log X)^2}{\epsilon_Y^2
+ b^2 \epsilon_X^2 }
\end{equation}
for each of the different datasets $i$ considered in our sample.
In the equation above, $\epsilon_Y = \sigma_L / L_{\rm X}$ and $\epsilon_X
= \sigma_T / T_{\rm gas}$.
We assume a 10 per cent uncertainty on the luminosity measurements
(i.e., $\epsilon_Y = 0.1$) and $\sigma_T$ equal to the maximum of the
68.3 per cent confidence asymmetric errors on the temperature.

In particular, combining several different datasets, we follow 
the indications in Lahav et al. (2000) for minimizing a ``weighted"
merit function 
\begin{equation}
C = \sum_i N_i \ln \left( \chi^2(a,b)_i \right),
\end{equation}
where $N_i$ is the number of data present in the dataset $i$.
This formalism allows us to consider some effective Hyper-Parameters,
$hp = N_i/\chi^2(a,b)_i$, that provide a diagnostic 
on the systematic effects (more relevant for low $hp$)
of that specific dataset in the 
overall modeling. Considering this, the \beppo data are
better modelled than the other two datasets ($hp \sim$ 3
vs. 0.2). 

The errors on the best-fit parameters are estimated from 500 bootstrap
replications of the fit. In Table~3, we show both the best-fit 
parameters $a$ and $b$ and the values that characterize their 
distribution after 500 bootstraps (i.e., the median and the 
16 and 84 percentile that corresponds to the lower and upper
1 $\sigma$ level in a Gaussian distribution).
%% average and dispersion around the mean).

Here we note that the computation of the physical properties presented
in the present work follows the technique used in Allen (2000) and
differs from the analysis done by Markevitch in the fact that we
try to model spectrally the emission from the core, whereas
Markevitch cuts the core emission from the spectral analysis
and extrapolates the contribution from the core to the total
luminosity from the external emissivity.
To take into account the different way in which the bolometric
luminosity is estimated, we leave free the normalization, $a$,
in the $L-T$ relation in eqn.~\ref{eq:fit} when we perform a linear
fit in each dataset. Only the slope parameter is linked
among the three datasets.

Combining the three datasets, we observe a slope in the 
$L-T$ relation of about 2.7. From the distribution of the
best-fit values after bootstrap resampling, we are still 
$\sim 4 \sigma$ far (too steep) from the value of 2 predicted
from the simple gravitational collapse of plasma in a cluster dark matter. 

This result is consistent, within $1 \sigma$, among the three different
datasets here considered and with the Bayesian analysis applied to the 
Allen (2000)'s and Markevitch (1998)'s sample by Reichart, Castander \&
Nichol (1999). 

\begin{table*}
\caption[]{Best-fit results on the luminosity--temperature relation
for the data in our sample with median values of 
$L =30.0 \times 10^{44}$ erg s$^{-1}$ and $T = 8.0$ keV.
The first two columns state the sample analysed and the number of
objects in exam. The third column indicates the number of objects 
considered in each subsample and the measured $\chi^2$ for the 
global best-fit. The scatter in the fourth column is defined
as $[ \sum( \log Y -\log a -b \log X )^2/N ]^{1/2}$,
where $Y = L_{\rm X} / (10^{44} h_{50}^{-2}$ erg s$^{-1})$
and $X = T_{\rm gas} / (10 {\rm keV})$.
Note that the scatter along the X-axis can be estimated as
$\sigma_{\ln X} = \sigma_{\ln Y} / b$.
In the last column, we quote the best-fit slope after applying the 
technique described in Section~4, and the median and $1 \sigma$ error
after 500 bootstrap replications.
}
\begin{tabular}{c@{\hspace{.6em}} c@{\hspace{.6em}} cccc}
cut & N data & N sample ($\chi^2$) & scatter & & $b$  \\
 \\
only \beppo data & 3 & 3 (0.9) & 0.327 & & $2.21 \ (1.50 \pm 1.13)$\\
only Allen (2000) & 30 & 30 (141.3) & 0.512 & & $2.41 \ (2.41 \pm 0.35)$\\
only Markevitch (2000) & 31 & 31 (108.7) & 0.240 & & $2.78 \ (2.78 \pm 0.16)$\\
all & 64 & 3 (1.0), 30 (146.8), 31 (109.9) & 0.437 & &
$2.69 \ (2.76 \pm 0.17)$ \\
 \\
$L> 10 \times 10^{44}$ erg s$^{-1}$ & 53 & 3 (1.0), 30 (141.5), 20 (71.0) & 
0.420 & & $2.45 \ (2.57 \pm 0.29)$ \\
 \\
$T> 6$ keV & 46 & 3 (1.2), 25 (86.0), 18 (61.6) & 0.388 & &
$2.72 \ (2.87 \pm 0.30)$ \\
\end{tabular}
\end{table*}

\section{DISCUSSION AND CONCLUSIONS}

From the \beppo observation, we can constrain the ambient temperature
in the clusters of galaxies RXJ1347-1145, Z3146 and A2390 with a 
relative uncertainty of (17/41), (11/41), (7/9)
per cent in the (lower/upper) boundary 
(at the 90 per cent confidence level) respectively, when a 
cooling flow model is included in the spectral analysis.

Although the instrument characteristics of \beppo do not allow us
to constrain firmly any absorption intrinsic to the cluster atmosphere 
and/or the normalization of the emission from the central cooling gas,
the measurements of the ambient gas temperature 
are in agreement, within the 90 per cent confidence level,
with the results from the analysis of \asca data in Allen (2000). 
In particular, the plasma temperature in RXJ1347-1145 appears
well defined between 13.2 and 22.3 keV (90 per cent confidence level)
and above 12.1 keV at the $3 \sigma$ level.
This measurement of the temperature is in good agreement with an 
independent estimate of 16.2 ($\pm$ 3.8, $1 \sigma$ level) keV 
obtained from the observed X-ray gas density and the 
measured Sunyaev--Zeldovich (1972) 
distorsion of the cosmic microwave background in the direction
of RXJ1347-1145 (Pointecouteau et al. 1999; see also Komatsu et al. 1999).

\begin{figure}
\hspace*{-1.cm}
\psfig{figure=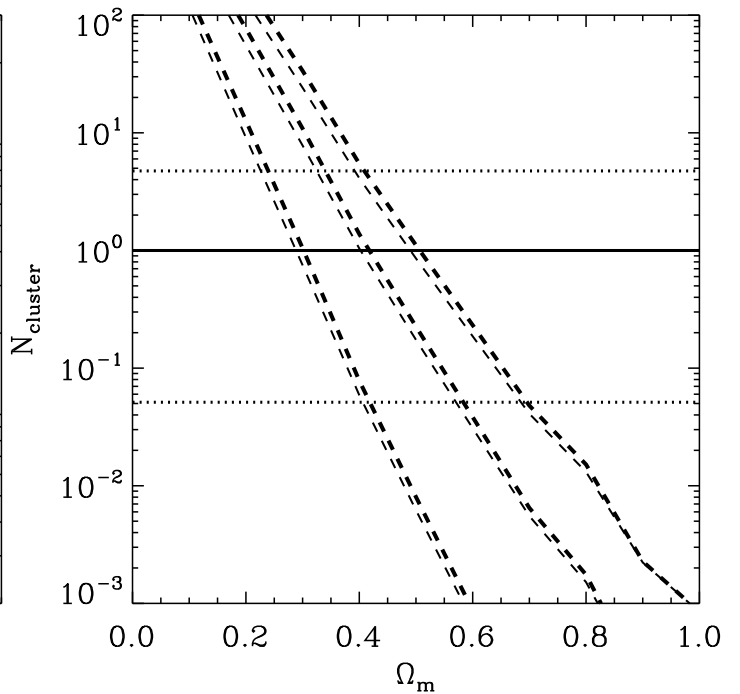,width=.6\textwidth}
\caption[]{The expected number density of clusters with gas temperature 
larger than 13.2, 15.9, 22.3 keV (from right to left in the figure, 
respectively; cf. Table~2 for the constraint on the gas temperature
of RXJ1347-1145) in the redshift interval $[0.4,0.5]$. 
The dashed lines represent the distribution in a ``$\Omega_{\rm m}
+\Omega_{\Lambda}=1$" Universe (thick) and in a open Universe
(thin). The error bars, around the observed value of 1 cluster
with the given characteristics, are at the 90 per cent level
of confidence, according to the 
Poisson single-sided distribution tabulated in Gehrels (1986).
} \label{fig:cosmo} \end{figure}

Following the arguments of many authors (e.g. Bahcall \& Fan 1998, 
Donahue et al. 1999, Henry 2000) the existence of such a hot
cluster at redshift 0.45 is highly unlikely in an Einstein -- de Sitter
universe. We can estimate this probability, calculating the 
expected evolution of collapsed structures with a given minimum mass 
according to the Press-Schechter (1974) formalism.
Integrating the number of collapsed objects with gravitating mass above
that corresponding to a temperature of 15.9 
keV over the redshift range of $0.4-0.5$,
and requiring {\it at least} the existence of RXJ1347-1145,
the single detection provides a stringent upper limit of 
$\Omega_{\rm m} \la$ 0.43. Considering 
also the lower and upper end of the range of the acceptable
gas temperature of 13.2 and 22.3 keV,
with the 90 per cent uncertainty on the single detection, 
we constrain the cosmological parameter $\Omega_{\rm m}$
between 0.25 and 0.70 (cf. Fig.~\ref{fig:cosmo}).

Future {\it Chandra} and {\it XMM} observations of these highly luminous
clusters of galaxies at intermediate redshift will resolve spatially the
profiles of the gas temperature, abundance and deposition rate. 

\section*{ACKNOWLEDGEMENTS} 
This paper has made use of linearized event files produced at the 
\beppo Science Data Center.
We thank T.~Oosterbroek and A.~Parmar for their help in using the {\sc SAXDAS}
routine {\it lemat}. The referee, A.~Edge, is thanked for his 
suggestions in improving the presentation of this work.
We acknowledge the support of the Royal Society. 
% S. Molendi is warmly thanked for his help in using the {\sc SAXDAS} 
% routine, {\it effarea}.

\end{document}